\documentclass[aps,prd,superscriptaddress,long,notitlepage,balancelastpage,nofootinbib,floatfix,twocolumn]{revtex4-1}
\pdfoutput=1

\usepackage{amsmath,mathtools,amssymb,amsthm,amsxtra,overpic,bbm,epsfig,subfigure,url,bm}
\usepackage{hyperref}
\usepackage{mathrsfs}
\usepackage{color,xcolor}
\usepackage{comment}
\usepackage{float}
\usepackage{enumitem}
\usepackage{slashed}
\usepackage{multirow}
\usepackage[normalem]{ulem}
\usepackage{braket}

\definecolor{nicered}{rgb}{0.5,0.,0.}
\definecolor{nicegreen}{rgb}{0.,0.5,0.}
\definecolor{niceblue}{rgb}{0.,0.,0.5}
\hypersetup{
	colorlinks=true,
	linkcolor=black,
	filecolor=nicegreen,      
	urlcolor=niceblue,
	citecolor=nicered,
}

\makeatletter
\newcommand*{\balancecolsandclearpage}{%
	\close@column@grid
	\cleardoublepage
	\twocolumngrid
}
\makeatother
\begin{document}
\title{Flavor composition of supernova neutrinos}

\author{\bf Antonio Capanema}
\email[E-mail:]{antoniogalvao@aluno.puc-rio.br}
\affiliation{Departamento de Física, Pontifícia Universidade Católica do Rio de Janeiro, Rio de Janeiro 22452-970, Brazil}

\author{\bf Yago Porto}
\email[E-mail:]{yago.porto@ufabc.edu.br}
\affiliation{Centro de Ciências Naturais e Humanas, Universidade Federal do ABC, 09210-170 Santo André, SP, Brazil}
\affiliation{Instituto de F{\'i}sica Gleb Wataghin, Universidade Estadual de Campinas, 13083-859 Campinas, SP, Brazil}

\author{\bf Maria Manuela Saez}
\email[E-mail:]{manuela.saez@berkeley.edu}
\affiliation{RIKEN Interdisciplinary Theoretical and Mathematical Sciences Program (iTHEMS), 2-1 Hirosawa, Wako, Saitama
351-0198, Japan}
\affiliation{Department of Physics, University of California, Berkeley, California 94720, USA}

\begin{abstract}
Predicting the flavor composition of neutrinos from supernovae is a challenging task, primarily due to the high neutrino densities at their core. In such an environment, neutrino self-interactions give rise to collective effects that have dramatic yet poorly understood consequences for their flavor evolution. In this paper, however, we show that standard matter effects in the outer layers of supernovae can significantly constrain the flavor composition of the neutrino flux.
We assume that, since a large number of neutrinos undergo different evolutions within the core, their state upon entering the MSW-dominated region is affected by decoherence. This assumption simplifies the problem and suggests that the fraction of neutrinos with electron flavor reaching Earth, denoted as $f_{\nu_e}$, is constrained to be less than $0.5$ for all energies throughout the emission phase in the case of normal mass ordering. In contrast, for inverted mass ordering, we anticipate neutrinos arriving in near flavor equipartition ($f_{\nu_e} \approx 1/3$).
These predictions, and consequently their underlying assumptions, could be tested by future observations and may provide valuable insights into the properties of neutrino fluxes emerging from supernovae.
\noindent 
\end{abstract}
\maketitle
\textbf{{\textit {Introduction.---}}} During a core-collapse supernova (SN) event, the immense gravitational binding energy of the progenitor star is predominantly liberated through a rapid emission of neutrinos, typically lasting around 10~s \cite{Colgate:1966ax, Arnett:1966}. After being generated in the SN core, these neutrinos traverse the mantle and envelope, where they are thought to be pivotal in driving the explosion \cite{Wilson:1985, Bethe:1985sox}. 

\begin{figure*}[htb!]
    \includegraphics[width=\textwidth]{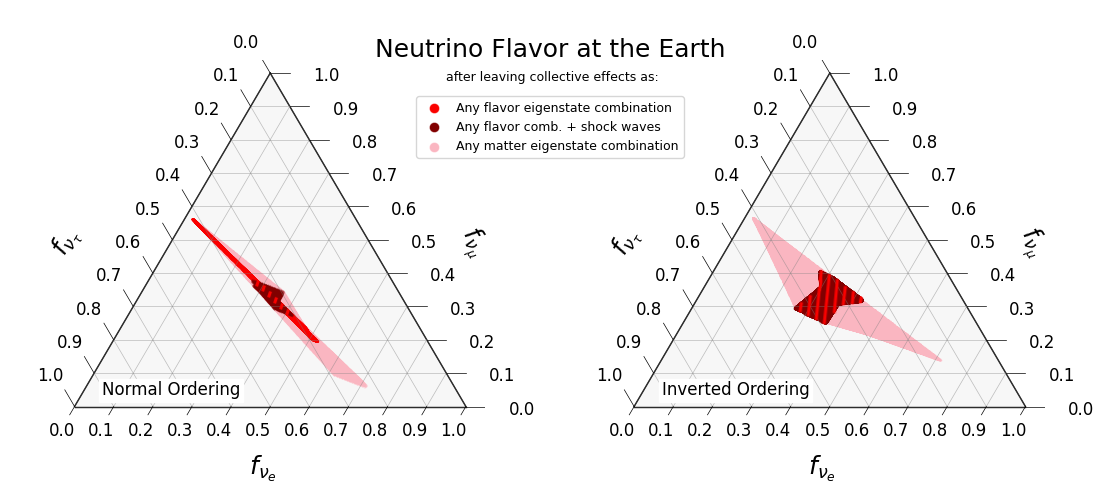}
    \caption{Allowed regions for the flavor composition of supernova neutrinos arriving at Earth for the cases of normal mass ordering (left) and inverted mass ordering (right). The colors represent three different scenarios. Red: Flux emerging from collective effects in an
    arbitrary flavor configuration (this can be approximated by an ensemble of flavor eigenstates; see text for details), and propagation in the outer layers is adiabatic, $f_{\nu_e}^{NO} \lesssim 0.5$ and $f_{\nu_e}^{IO} \approx 1/3$.  Maroon: Flux emerging from collective effects in an arbitrary flavor configuration, but nonadiabatic propagation through the $H$-resonance (presence of shock wave), $f_{\nu_e}^{NO} \approx f_{\nu_e}^{IO} \approx 1/3$. Stripes show overlaps between red and maroon regions. Pink: Flux of matter eigenstates in arbitrary amounts emerging from collective effects, $f_{\nu_e}^{NO} \approx f_{\nu_e}^{IO} \lesssim 0.7$. In this figure, we consider the $3 \sigma$ range of variability only for the parameters $\theta_{12}$ and $\theta_{13}$, which affect the fraction of electron flavor $f_{\nu_e}$.}
    \label{fig1}
\end{figure*}

In 1987, the detection of a few tens of SN neutrinos coming from the Large Magellanic Cloud confirmed the basic features of the core-collapse models, such as the mean energy and the duration of the neutrino burst \cite{Kamiokande-II:1987idp, Bionta:1987qt, Alekseev:1988gp} (see \cite{Fiorillo:2023frv} for a recent analysis). Nevertheless, the intricacies of the SN physics remain a challenge, largely due to the uncertainties surrounding neutrino flavor transformations in the core \cite{Tamborra:2020cul, Ehring:2023lcd, Ehring:2023abs, Nagakura:2023mhr}. In such a neutrino-dense environment, the coherent forward scattering of neutrinos onto each other gives rise to collective effects that drive a nonlinear flavor evolution \cite{Pantaleone:1992eq, Pantaleone:1992xh, Duan:2010bg, Mirizzi:2015eza, Chakraborty:2016yeg}. These self-induced conversions are referred to as ``slow" or ``fast" depending on the characteristic spatial scale in which they manifest. Slow self-induced conversions develop over a length scale governed by the vacuum oscillation frequency, which is on the order of a few kilometers for typical SN neutrino energies ($\sim 10$ MeV) \cite{Duan:2010bg, Mirizzi:2015eza}. These conversions take place in an intermediate zone situated $\sim 100$ km away from the neutrino decoupling region, but before the standard matter-induced resonant neutrino conversion occurs \cite{Wolfenstein:1977ue, Mikheyev:1985zog, Mikheev:1986wj, Mikheev:1987jp}.
Conversely, fast self-induced conversions have characteristic scales as small as centimeters and could significantly influence the flavor evolution, energy spectrum, and angular distribution of each neutrino species near the decoupling region \cite{Sawyer:2005jk, Sawyer:2008zs, Sawyer:2015dsa, Chakraborty:2016lct, Izaguirre:2016gsx}. In addition, it was recently found that coherent flavor conversion can also be induced by incoherent collisions in the SN environment \cite{Johns:2021qby}. The lack of a full solution to the quantum kinetic equations governing this collective behavior leads to unpredictable consequences for the SN dynamics \cite{Ehring:2023abs, Nagakura:2023mhr} and, ultimately, for the neutrino signal detected at Earth \cite{Abbar:2024nhz}. 

Amidst this conundrum, it is natural to wonder if a meaningful improvement in our understanding of the core-collapse phenomenon and neutrino propagation will be possible, even after the observation of the next galactic SN. In this paper, we show that many features of the neutrino flux coming to Earth can be understood by making a simple assumption regarding the state of the neutrino ensemble as they emerge from the SN core. Namely, due to the large number of neutrinos interfering randomly, we expect coherence to be lost, rendering the density matrix for the ensemble diagonal in some basis (which need not be the flavor basis or any specific basis \textit{a priori}) as they enter the MSW-dominated region. We then show that, for most of these bases, matter effects alone impose significant constraints on the flavor composition of supernova neutrinos.
Assuming normal ordering (NO) of neutrino masses, the fraction of electron neutrinos arriving at Earth, denoted by $f_{\nu_e}^{NO}$, is found to remain consistently below $0.5$ for all energies during the emission phase, and gradually converges to near flavor equipartition ($f_{\nu_e}^{NO} \approx 1/3$)\footnote{For our purpose, ``flavor equipartition'' simply means $f_{\nu_e}=1/3$, since discrimination between $\nu_\mu$ and $\nu_\tau$ signals in current and next-generation detectors is challenging \cite{Akhmedov:2002zj}.} during the passage of shock waves. In the inverted ordering (IO) scenario, we predict flavor equipartition ($f_{\nu_e}^{IO} \approx 1/3$) as a generic outcome, which remains robust throughout the entire neutrino burst, even in the presence of shock waves. 

If confirmed by the next SN observation, these results will not only validate our fundamental understanding of the SN dynamics and neutrino conversion in the outermost layers, but also elucidate crucial properties of the neutrino fluxes emerging from the region of self-induced conversions.

\vspace{0.05in}

\textbf{{\textit {Standard Matter Effects.---}}} Neutrino-neutrino interactions are expected to play a subdominant role in flavor evolution after the first few hundred kilometers from the neutrino decoupling region \cite{Tamborra:2020cul, Mirizzi:2015eza}. At the outermost layers, neutrino evolution is predominantly affected by the matter potential generated by the coherent forward scattering of neutrinos on electrons, protons and neutrons \cite{Wolfenstein:1977ue}. In practice, however, only the charged current interactions between electrons and electron (anti)neutrinos influence neutrino oscillations. Consequently, the effective Hamiltonian in the outer SN regions can be written in the flavor basis as follows
\begin{equation} \label{H}
    \mathcal{H}_{\nu}=\frac{1}{2 E} \,U\left(\begin{array}{ccc}0 & 0 & 0 \\ 0 & \Delta m_{21}^2 & 0 \\ 0 & 0 & \Delta m_{31}^2\end{array}\right) U^{\dagger}+\left(\begin{array}{ccc}V_e & 0 & 0 \\ 0 & 0 & 0 \\ 0 & 0 & 0\end{array}\right).
\end{equation}
Here, $U$ represents the Pontecorvo–Maki–Nakagawa– Sakata (PMNS) matrix, $\Delta m^2_{ij}$ stands for the mass-squared differences and $E$ denotes the neutrino energy. The matter potential affects only the $\nu_e$ flavor and has the form $V_e= \sqrt{2} G_F  \rho_m Y_e/m_N$, where $G_F$ is the Fermi constant, $Y_e$ is the electron number fraction, $m_N$ is the nucleon mass and $\rho_m$ is the local matter density. It is important to consider the implicit dependence of $V_e$ and $\mathcal{H}_\nu$ on time and space, as each neutrino follows a different trajectory inside the SN, and the physical conditions along each trajectory change with time. For antineutrinos, the matter potential has the opposite sign, $\Bar{V}_e=-V_e$. Thus, the equation of motion for neutrinos in the SN outer layers is given by
\begin{equation} \label{eq-motion}
    i \frac{d \rho_\nu}{d t} = [\mathcal{H}_\nu,\rho_\nu],
\end{equation}
where $\rho_\nu$ is the density matrix describing the neutrino flavor state. Observe that Eq.~\eqref{eq-motion} is a simplified version of the complete expression that takes into account neutrino-neutrino interactions, advection, and collisions \cite{Sigl:1993ctk, Cardall:2007zw,Fiorillo:2024wej,Fiorillo:2024fnl}.

As neutrinos propagate toward vacuum and the matter density ($\rho_m$) gradually decreases, they encounter two resonance layers where effective flavor conversions can occur \cite{Mikheyev:1985zog, Mikheev:1986wj, Mikheev:1987jp}. These so-called $H$- and $L$-resonances occur when the following conditions are satisfied:
\begin{equation} \label{resonance-conditions}
  \sqrt{2} G_F \frac{Y_e}{m_N} \rho_m^{ \rm res}=  \frac{\Delta m^2_{i1} \cos 2 \theta_{1i}}{ 2  E},
\end{equation}
where ${\rm res}=H(L)$ corresponds to $i=3(2)$. For efficient flavor conversion, adiabaticity must be satisfied at the resonance layers. The degree of adiabaticity can be estimated by considering the parameter
\begin{equation} \label{adiabaticity}
    \gamma = \frac{\Delta m^2_{i1} \sin^2 2 \theta_{1i}}{2E \cos 2 \theta_{1i}} \left( \frac{1}{Y_e \rho_m^{\rm res}} \left| \frac{d (Y_e \rho_m) }{d r} \right|_{\rm res} \right)^{-1},
\end{equation}
and propagation is adiabatic for $\gamma > 1$ \cite{Dighe:1999bi}.

At densities above $\rho_m^H$, $V_e$ dominates in Eq.~\eqref{H} and suppresses vacuum mixing. As a result, $\nu_e$ effectively decouples from the other flavors and behaves as the heaviest eigenstate in matter, which corresponds to $\nu_3^m$ in NO and $\nu_2^m$ in IO, where the superscript $m$ denotes matter eigenstates \cite{Dighe:1999bi}. Conversely, the other flavors are equally affected by matter\footnote{Differences between $\nu_\mu$ and $\nu_\tau$ are negligible at densities below $10^7$ $\text{g/cm}^3$ \cite{Botella:1986wy}.} and oscillate among themselves with frequency given by the vacuum terms and near maximal mixing ($\theta_{23} \approx 49^\circ$). Therefore, the eigenstates of the $\nu_\mu - \nu_\tau$ subspace at $\rho_m \gg \rho_m^H$ are given by approximately equal mixtures of $\nu_\mu$ and $\nu_\tau$. In NO, these matter eigenstates are $\nu_1^m$ and $\nu_2^m$, while in IO, they are $\nu_1^m$ and $\nu_3^m$.

\vspace{0.05in}

{\textbf {\textit {Neutrino state emerging from self-induced conversions.---}}} Assuming an emission of $10^{58}$ neutrinos over the $10$~s duration of a galactic SN burst \cite{SajjadAthar:2021prg}, at a distance of 10 kpc, we estimate the flux reaching Earth to be $N \sim 10^{15}$ neutrinos per square meter per second. 
Before traveling to Earth, these neutrinos reach the boundary between the region of self-induced conversions and the region of standard matter effects, where any superposition of electron neutrinos with other flavors, initially formed by self-induced conversions, becomes negligible. Therefore, at this point, the density matrix for the ensemble of $N$ neutrinos in the flavor basis $F$ is
\begin{equation} \label{rho-nu}
    \rho_{\nu}^{F}= \frac{1}{N} \sum_{k=1}^N \left(\begin{array}{ccc}\alpha_{k}^2 & 0 & 0 \\ 0 & \beta_{k}^2 & \beta_{k} \gamma_{k} e^{-i \phi_k} \\ 0 & \beta_{k} \gamma_{k} e^{i \phi_k} & \gamma_{k}^2\end{array}\right),
\end{equation}
where $\alpha_k^2$ represents the probability of the $k$th neutrino to interact as $\nu_e$, $\beta_k^2$ as $\nu_\mu$, and $\gamma_k^2$ as $\nu_\tau$, with the condition that $\alpha_k^2 + \beta_k^2 + \gamma_k^2 = 1$. The off-diagonal terms in Eq.~\eqref{rho-nu} represent the interference between the $\nu_\mu$ and $\nu_\tau$ states in matter. Here, we propose that, since the neutrinos reaching Earth may have been produced at completely different locations in the SN core, and have followed a wide variety of trajectories and evolutions before reaching the region of standard matter effects, the sum of their interference terms will destructively interfere and vanish in some basis $B$. In the appendix, we show that even if destructive interference occurs in a basis other than the flavor basis, the off-diagonal terms in the flavor basis are suppressed.\footnote{There is no reason to assume that there is a preferred basis in which the complex contributions to the off-diagonal of Eq.~\ref{rho-nu} average to zero, nor that this basis remains the same at different times during the emission.} Based on this assumption, we can approximate the general state in Eq.~\eqref{rho-nu} by an ensemble of flavor eigenstates with arbitrary proportions (i.e., $\rho_\nu^F$ is approximately diagonal) and narrow down the range of possible values for $f_{\nu_e}$ in the following sections. The reader should keep in mind that, if $B\neq F$, incomplete destructive interference in Eq.~\eqref{rho-nu} is still possible. We treat this as theoretical uncertainties in our results in the upcoming sections (see the appendix for details).

\vspace{0.05in}


{\textbf {\textit {Results: Flavor ratios at the Earth.---}}} In the following discussion, we show that under the assumption that interference terms are suppressed in Eq.~\eqref{rho-nu} and that neutrino propagation is adiabatic, the $\nu_e$ fraction at Earth is given by $f_{\nu_e}^{NO} \lesssim 0.5$ and $f_{\nu_e}^{IO} \approx 1/3$. These results correspond to the red-shaded regions in Fig.~\ref{fig1}.

To better understand the emergence of these constraints, we characterize the evolution within the zone of standard matter effects in terms of flavor ratios $(f_{\nu_e},f_{\nu_\mu},f_{\nu_\tau})$. Assuming NO and adiabatic propagation, it follows that an initially produced $\nu_e$ arrives at Earth as $\nu_3$ \cite{Dev:2023znd}:
\begin{equation} \label{NO1}
(1,0,0)_{SN}  \rightarrow\left(\left|U_{e 3}\right|^2,\left|U_{\mu 3}\right|^2,\left|U_{\tau 3}\right|^2\right)_{\oplus}.
\end{equation}
Conversely, either of the nonelectron flavors $\nu_\mu$ or $\nu_\tau$ arrive at the Earth approximately as equal mixtures of $\nu_1$ and $\nu_2$. Therefore,
\begin{align} \label{similar-numbers}
&\hspace{0.5 cm} \hspace{0.5 cm} \hspace{0.5 cm}(0,1,0)_{SN} \hspace{0.5 cm} \text{or} \hspace{0.5 cm}  (0,0,1)_{SN}\rightarrow \nonumber  \\
&\frac{1}{2}\left(\left|U_{e 1}\right|^2+\left|U_{e 2}\right|^2,\left|U_{\mu 1}\right|^2+\left|U_{\mu 2}\right|^2,\left|U_{\tau 1}\right|^2+\left|U_{\tau 2}\right|^2 \right)_{\oplus}.
\end{align}
For any initial combination $(a,b,c)_{SN}$,\footnote{In the context of Eq.~\eqref{rho-nu}, we have $a= \frac{1}{N} \sum_k \alpha_k^2$, $b=\frac{1}{N}\sum_k \beta_k^2$, and $c=\frac{1}{N}\sum_k \gamma_k^2$.} we obtain on Earth the $\nu_e$ fraction
\begin{equation} \label{NO2}
    f_{\nu_e}^{NO}= a \left|U_{e 3}\right|^2 + b \frac{\left|U_{e 1}\right|^2+\left|U_{e 2}\right|^2}{2} + c \frac{\left|U_{e 1}\right|^2+\left|U_{e 2}\right|^2}{2}.
\end{equation}
Because of the unitarity of the PMNS matrix and $a+b+c=1$, Eq.~\eqref{NO2} simplifies to 
\begin{equation} \label{NO4}
    f_{\nu_e}^{NO}= \frac{1}{2} \left( 1 - \left|U_{e 3}\right|^2    \right) + \frac{a}{2} \left( 3 \left|U_{e 3}\right|^2 -1   \right).
\end{equation}
Adopting $\left|U_{e 3}\right|^2 \approx 0.02 \ll 1$ \cite{deSalas:2020pgw}, we obtain
\begin{equation} \label{NO5}
    f_{\nu_e}^{NO} \approx \frac{(1-a)}{2} \lesssim 0.5. 
\end{equation}
The theoretical uncertainty on this upper limit is $\sigma_{\nu_e}^{NO} \in [-0.08,0.06]$ (see the appendix for details).
In the IO scenario, the argument mirrors the one between Eq.~\eqref{NO1} and \eqref{NO4}, with the interchange of the PMNS matrix elements $U_{\alpha 3}$ and $ U_{\alpha 2}$ for any flavor $\alpha$. Thus, by replacing $U_{e 3}$ with $U_{e 2}$ in Eq.~\eqref{NO4}, we have
\begin{equation} \label{IO4}
    f_{\nu_e}^{IO}= \frac{1}{2} \left( 1 - \left|U_{e 2}\right|^2    \right) + \frac{a}{2} \left( 3 \left|U_{e 2}\right|^2 -1   \right).
\end{equation}
Assuming $\left|U_{e 2}\right|^2 \approx 1/3$ \cite{deSalas:2020pgw}, we obtain\footnote{According to Eqs.~\ref{IO4} and \ref{IO5}, any initial flavor configuration in the solar core leads to $f_{\nu_e} \approx 1/3$ at the Earth for both mass orderings.}
\begin{equation} \label{IO5}
    f_{\nu_e}^{IO} \approx \frac{1}{3},
\end{equation}
with uncertainty $\sigma_{\nu_e}^{IO} \in [-0.1,0.1]$.
The red-shaded regions in Fig.~\ref{fig1} correspond to $f_{\nu_e}^{NO}$ and $f_{\nu_e}^{IO}$ after varying $\theta_{12}$ and $\theta_{13}$ within their $3 \sigma$ allowed range and using the best-fit values for the other oscillation parameters (the impact of $\theta_{23}$ and $\delta_{CP}$ on $f_{\nu_e}$ range is less than $10\%$) \cite{deSalas:2020pgw}.

The results in Eq.~\eqref{NO5} and \eqref{IO5} are applicable to neutrinos of any energy and at any instant in time. Nonetheless, exceptions may be possible when a shock wave is present, leading to modification of Eq.~\eqref{NO5} at certain parts of the energy spectrum.

\vspace{0.05in}


{\textbf {\textit {Results in the presence of shock waves.---}}} When adiabatic propagation cannot be guaranteed, transitions between different matter eigenstates may occur at the $H$- and $L$-resonance layers. Furthermore, if the matter profile is affected by shock waves, multiple resonances of the $H$- and $L$-types could appear \cite{Schirato:2002tg, Tomas:2004gr}. If an even number of $H$-resonances are nonadiabatic, results do not change [neglecting phase effects \cite{Dasgupta:2005wn}]. However, if an odd number of the $H$-resonances are nonadiabatic, the outcome is modified in the NO scenario once transitions $\nu_3^m \leftrightarrow \nu_2^m$ that happen at one of the resonance layers are not reverted by another transition of the same type. It is straightforward to determine the consequence of an odd number of nonadiabatic $H$-resonances: transitions of the type $\nu_3^m \leftrightarrow \nu_2^m$ imply that matrix elements $U_{\alpha 3}$ and $ U_{\alpha 2}$ should be interchanged in Eqs.~\eqref{NO1}-\eqref{NO4}. The argument is similar to the one that leads to Eqs.~\eqref{IO4} and \eqref{IO5}, resulting in
\begin{equation}
    f_{\nu_e}^{NO} (\text{non-ad $H$}) \approx \frac{1}{3}.
\end{equation}
Because the position of the $H$-resonance is energy dependent, the portion of the energy spectrum influenced by the transient nonadiabaticity provides information about the location of the shock wave front \cite{Schirato:2002tg, Tomas:2004gr}.

In the IO scenario, the $H$-resonance does not occur in the neutrino channel. Therefore, the result is identical to the adiabatic case:
\begin{equation}
    f_{\nu_e}^{IO} (\text{non-ad $H$}) \approx \frac{1}{3}.
\end{equation}

The impact of the nonadiabatic $H$-resonance in neutrino propagation is illustrated by the maroon-striped region in Fig.~\ref{fig1}. Once again, we vary only $\theta_{12}$ and $\theta_{13}$ within their 3$\sigma$ uncertainties. We conclude that the presence of shock wave fronts at the $H$-resonances causes the neutrino system to converge to approximate flavor equipartition at Earth for both mass orderings.

Although shock wave fronts can also penetrate the $L$-resonance layer, adiabaticity is never strongly violated there \cite{Mirizzi:2015eza}. Even in cases where such violations may happen, they typically occur at later times (beyond $10$~s post-bounce) \cite{Tomas:2004gr}, so we do not consider them here.

\vspace{0.05in}


{\textbf {\textit {Additional remarks.---}}} The extreme values of the interval in Eq.~\eqref{NO5} suggest that neutrinos would emerge from the inner regions either entirely in the electron flavor ($f_{\nu_e}^{NO} \approx 0$) or completely devoid of it ($f_{\nu_e}^{NO} \approx 0.5$). These scenarios are probably unlikely, as they would require a specific mechanism to produce neutrinos in these extreme states.\footnote{Exception is the early stage of emission \cite{Serpico:2011ir}, during which self-induced conversions are negligible and neutrinos emerge mostly as $\nu_e$.} Therefore, the $f_{\nu_e}^{NO}$ range should often approach the one for IO [Eq.~\eqref{IO5}].

The results above emerge from the maximal mixing between $\nu_\mu$ and $\nu_\tau$ in matter, which restricts two of the three mass eigenstates to reach Earth in similar numbers [Eq.~\eqref{similar-numbers}]. It is interesting to contrast the results above with the situation where any combination of mass eigenstates is allowed, $f_{\nu_e} \lesssim 0.7$ \cite{Horiuchi:2018ofe}, given by the pink region in Fig.~\ref{fig1}.

\vspace{0.05in}


{\textbf {\textit {Conclusion.---}}} We have shown that the flavor composition of SN neutrinos reaching Earth can be constrained due to the distinct flavor evolution of neutrinos in the core, which leads to the decoherence of the neutrino ensemble upon reaching the MSW-dominated outer regions. For NO, we predict that the proportion of $\nu_e$ is always smaller than $0.5$, whereas for IO, we expect near flavor equipartition ($f_{\nu_e} \approx 1/3$).

Ultimately, this article advocates for a phenomenological approach to comprehensively address the intricacies of neutrino flavor conversion in SN environments. Experimental challenges along the way include distinguishing between $\nu_x$ and $\Bar{\nu}_x$ in neutral current scatterings \cite{Jachowicz:2004we}. This approach can also be extended to accommodate non-standard scenarios, potentially unveiling novel phenomena underlying the physics of neutrino masses and mixing \cite{Jana:2022tsa, Jana:2023ufy}.


\vspace{0.09in}
{\textbf {\textit {Acknowledgments.---}}} We thank Orlando Peres for useful discussions. A.C.
thanks for the support received by the scholarships CAPES/PROEX No. 88887.511843/2020-00, CNPq No. 140316/2021-3, CAPES-PrInt No. 88887.717489/2022-00 and FAPERJ No. E-26/204.138/2022. The work of Y.P.
was supported by the S\~{a}o Paulo Research Foundation (FAPESP) Grants No. 2023/10734-3 and No. 2023/01467-1, and by the National Council for Scientific and Technological Development (CNPq) Grant No. 151168/2023-7. M.M.S. is thankful for the support provided by iTHEMS and ABBL from RIKEN, the N3AS Physics Frontier Center, and the NSF under cooperative Agreement No. 2020275. This work would not have been possible without the collaborative environment provided by the Nordic Winter School on Multimessenger Astrophysics. We thank the event organizers, as well as the financial support by Markus Ahlers for A.C. to attend the event.

\vspace{0.09in}

\section*{Appendix}

{\textbf {\textit {Vanishing off-diagonal terms in an arbitrary basis.---}}}
\begin{enumerate}
    \item \textbf{Connection between flavor and matter basis}
    
To describe the evolution of neutrinos in the region of standard matter effects, we need to determine the fraction of each matter eigenstate ($\nu_1^m$, $\nu_2^m$, and $\nu_3^m$) entering this region. To go from the flavor basis to the matter basis, we rotate the subspace $\mu$-$\tau$ in $\rho_\nu^F$ [Eq.~\eqref{rho-nu}] by $\theta_{23} \approx 45^o$. For NO, the fractions are given by
\begin{equation} \tag{A1}
    f_{\nu_3^m}^{NO} = \frac{1}{N} \sum_k \alpha_k^2
\end{equation}
and
\begin{equation} \label{eq1:nu1-nu2} \tag{A2}
    f_{\nu_1^m(\nu_2^m)}^{NO} = \frac{1}{N} \sum_k^N \left( \frac{\beta_k^2 +  \gamma_k^2}{2} \pm \Re \left\{ \beta_k\gamma_k e^{i \phi_k} \right\} \right),
\end{equation}
where $\alpha_k$, $\beta_k$, $\gamma_k$ and $\phi_k$ are given in Eq.~\eqref{rho-nu} of the main text. Note that, due to the $\nu_\mu - \nu_\tau$ mixing in matter, the difference between the fractions of $\nu_1^m$ and $\nu_2^m$ is related to the off-diagonal elements of Eq.~\eqref{rho-nu} as 
\begin{equation} \tag{A3} \label{eq:difference}
    \left| f_{\nu_1^m}^{NO}-f_{\nu_2^m}^{NO} \right| = \left| \frac{2}{N} \sum_k \Re \left\{ \beta_k\gamma_k e^{i \phi_k} \right\} \right|
\end{equation}
(for IO, equations are the same, but with the interchange $\nu_3^m \leftrightarrow \nu_2^m$). Our hypothesis is that the off-diagonal term is suppressed, and two of the three matter eigenstates are present in similar proportions. A quantitative estimate of the values that the off-diagonal term is likely to assume can provide an indication of the extent to which our hypothesis holds (see section below).

\item \textbf{Vanishing interference terms in a basis other than the flavor basis}
    
    In Eq.~\eqref{rho-nu}, we describe the density matrix of neutrinos emerging from self-induced conversions in the flavor basis. As a result, when the off-diagonal elements of Eq.~\eqref{rho-nu} vanish, we can effectively describe the system as an ensemble of flavor eigenstates. Nevertheless, in the absence of a significant number of $\mu$ or $\tau$ leptons in the SN environment, there is nothing that distinguishes the flavor basis $F$ from any other basis $B$ in the $\mu$-$\tau$ sector. Therefore, we can rotate the $\mu$-$\tau$ sector to another arbitrary basis, 
\begin{equation} \label{rho-nu-prime} \tag{A4}
    \rho^{B}_{\nu}= \frac{1}{N} \sum_{k=1}^N \left(\begin{array}{ccc}\alpha_{k}^{2} & 0 & 0 \\ 0 & \beta_{k}^{\prime 2} & \beta_{k}^{\prime} \gamma_{k}^{\prime} e^{-i \phi^{\prime}_k} \\ 0 & \beta_{k}^{\prime} \gamma_{k}^{\prime} e^{i \phi^{\prime}_k} & \gamma^{\prime 2}_{k}\end{array}\right),
\end{equation}
and assume that the off-diagonal terms in this basis vanish:
\begin{equation} \label{vanish} \tag{A5}
    \frac{1}{N} \sum_{k=1}^N \beta^{\prime}_{k} \gamma^{\prime}_{k} e^{i \phi^{\prime}_k} \approx 0.
\end{equation}
In this case, after rotating to the flavor basis, we obtain
\begin{equation} \tag{A6} \label{rho-F-2}
    \rho_{\nu}^F= \frac{1}{N} \sum_{k=1}^N \left(\begin{array}{ccc}\alpha_{k}^2 & 0 & 0 \\ 0 & c_\theta^2 \beta_{k}^{\prime 2} + s_\theta^2 \gamma_k^{\prime 2} & \frac{1}{2} s_{2\theta} (\beta_k^{\prime 2}-\gamma_k^{\prime 2}) \\ 0 & \frac{1}{2} s_{2\theta} (\beta_k^{\prime 2}-\gamma_k^{\prime 2}) & s_\theta^2 \beta_{k}^{\prime 2} + c_\theta^2 \gamma_k^{\prime 2} 
    \end{array}\right),
\end{equation}
where $\theta$ parametrizes the rotation between the primed and flavor bases, with $s_\theta= \sin \theta$ and $c_\theta \equiv \cos \theta$. Here, we aim to demonstrate that even in such a situation, where destructive interference is known to occur in some other arbitrary basis $B$, the off-diagonal terms of $\rho_{\nu}^F$ are suppressed. Since the basis $B$ is arbitrary, we assume that $2\theta$ can assume any value in the range $[-\pi/2, \pi/2]$ ($s_{2 \theta}$ in the range $[-1,1]$). Moreover, $X \equiv \frac{1}{N} \sum_{k=1}^N (\beta_k^{\prime 2}-\gamma_k^{\prime 2})$ can take any value within the range $[-1,1]$ (in order to maximize our uncertainty, we consider $\sum\alpha_k^2 = 0$, concentrating all of the probability in the $\mu$-$\tau$ sector).

Thus, the question becomes: what is the probability that the absolute value of $\frac{1}{2} s_{2\theta} X$ exceeds a given value $Y$? 

To compute $P\left( \left|\frac{1}{2} s_{2\theta} X \right| > Y \right)$, we need to find $P\left( \frac{1}{2} s_{2\theta} X  > Y \right)$ and $P\left( \frac{1}{2} s_{2\theta} X  < - Y \right)$. In each instance, we need to separate the cases in which $X>0$ and $X<0$. Assuming $2\theta$ and $X$ are uniformly distributed,\footnote{This assumption is made with the sole purpose of simplifying calculations, as our argument is merely a proof of principle; This can be straightforwardly generalized to different choices of probability distributions for $\theta$ and $X$. On the other hand, assuming that neutrinos follow completely different evolutions, states given by $X = \pm 1$ (i.e., pure states) should be much less likely than other combinations that lead to mixed states. This more realistic scenario, which deviates from the uniform distribution, would make our results even more robust but would also be more complicated to handle. More comments on this will be provided later.} we obtain
\begin{align} 
    P\left( \frac{1}{2} s_{2\theta} X > Y \right)= \int_{2Y}^{1} \frac{dX}{2} \int_{\arcsin(2 Y/X)}^{\pi/2} \frac{d(2 \theta)}{\pi} \nonumber \\ + \int_{-1}^{-2Y} \frac{dX}{2}  \int_{-\pi/2}^{\arcsin(2 Y/X)} \frac{d(2 \theta)}{\pi} \nonumber \\ = \frac{1}{\pi} \int_{2Y}^{1} dX \left( \frac{\pi}{2}-\arcsin(2 Y/X) \right) \tag{A7},
\end{align}
where $dX/2$ is the probability of getting a number in the range between $X$ and $X+dX$ in a interval of total size $2$ ($X \in [-1,1]$), and $d(2 \theta)/\pi$ the probability of getting an angle in the range $2 \theta$ and $2 \theta + d(2 \theta)$ in an interval of total size $\pi$. The integration limits indicate that $X$ must be either greater than $2Y$ or smaller than $-2Y$ because otherwise no value of $s_{2\theta}$ satisfies the inequality. Following the same logic, we find
\begin{align} 
    P\left( \frac{1}{2} s_{2\theta} X < - Y \right)= \int_{2Y}^{1} \frac{dX}{2} \int^{-\arcsin(2 Y/X)}_{-\pi/2} \frac{d(2 \theta)}{\pi} \nonumber \\+ \int_{-1}^{-2Y} \frac{dX}{2}  \int^{\pi/2}_{-\arcsin(2 Y/X)} \frac{d(2 \theta)}{\pi} \nonumber \\ = \frac{1}{\pi} \int_{2Y}^{1} dX \left( \frac{\pi}{2}-\arcsin(2 Y/X) \right) \tag{A8}.
\end{align}
Therefore,
\begin{align} 
    P\left( \left| \frac{1}{2} s_{2\theta} X \right| > Y  \right)=P\left( \frac{1}{2} s_{2\theta} X > Y \right)+P\left( \frac{1}{2} s_{2\theta} X < - Y \right) \nonumber \\= \frac{2}{\pi} \int_{2Y}^{1} dX \left( \frac{\pi}{2}-\arcsin(2 Y/X) \right).\tag{A9}
\end{align}
Now we can calculate the probability that the off-diagonal term in $\rho_\nu^F$ is below a certain benchmark value $Y$, assuming that destructive interference occurred in the basis $B$. For instance, for $Y=0.2$, we find $P(\left| \frac{1}{2} s_{2\theta} X \right| <~ 0.2) = 1- P(>0.2) \approx 0.66$, implying that 66\% of the time (roughly $1$ standard deviation $\sigma$, in analogy to the normal distribution), the absolute value of the off-diagonal term is less than $0.2$ and peaks at zero (mean value). 

In light of Eq.~\eqref{eq:difference}, and assuming $\alpha_k^2 = 0$ for simplicity, we find that
\begin{align} 
    \left| f_{\nu_1^m}^{NO}-f_{\nu_2^m}^{NO} \right| = \left| 2 \times  \text{real part of off-diagonal term} \right| \nonumber \\ = \left| s_{2 \theta} X \right| \lesssim 0.4.\tag{A10}
\end{align}
These correspond to theoretical uncertainties in the calculated flavor ratios presented in the main text: we obtain $f_{\nu_e}^{NO} \lesssim 0.5 + \sigma_{\nu_e}^{NO}$ [Eq.~\eqref{NO5}] with $\sigma_{\nu_e}^{NO} \in [-0.08, 0.06]$, computed by varying $f_{\nu_1^m}^{NO} - f_{\nu_2^m}^{NO}$ within $[-0.4, 0.4]$. Additionally, we obtain $f_{\nu_e}^{IO} \approx 1/3 + \sigma_{\nu_e}^{IO}$ [Eq.~\eqref{IO5})] with $\sigma_{\nu_e}^{IO} \in [-0.1, 0.1]$ when varying $f_{\nu_1^m}^{IO} - f_{\nu_3^m}^{IO}$ within $[-0.4, 0.4]$.

Note, however, that these uncertainties are conservative, as $\sum \alpha_k^2$ generally tends to be nonzero. Therefore, the theoretical uncertainties should have less impact on the final results than expressed here, since all the values in the $\mu$-$\tau$ sector of $\rho_\nu^F$ (including off-diagonal elements) should be smaller by a factor of $(1 - \sum \alpha_k^2)$. Additionally, the distribution of $X$ may be non-uniform and peaked toward zero, further reinforcing our results and potentially narrowing the ranges for $f_{\nu_e}$ even more (we will return to this point in item~\ref{additional} below).

\item \textbf{Constrast with the situation of no mixing in the $\mu$-$\tau$ sector}

From the argument above, note that the suppression of the off-diagonal terms in the flavor basis$-$and consequently the small value of $\left| f_{\nu_1^m}^{NO} - f_{\nu_2^m}^{NO} \right|$ (likely $\lesssim 0.4$)$-$arises from the fact that, due to $\mu$-$\tau$ mixing in matter, there are many bases where decoherence can occur, and in each of these bases, there are certain parameter combinations for which the off-diagonal elements of Eq.~\eqref{rho-F-2} are small.
This should be contrasted with the situation where there is no $\mu - \tau$ mixing (for example, due to the presence of muons), and only the matter basis has physical significance:
\begin{equation} \tag{A11} 
    \rho_{\nu}^F=\rho_{\nu}^M= \frac{1}{N} \sum_{k=1}^N \left(\begin{array}{ccc}\alpha_{k}^2 & 0 & 0 \\ 0 &  \beta_{k}^{2} & 0 \\ 0 & 0 &  \gamma_k^{ 2} 
    \end{array}\right). \label{rho-M}
\end{equation}
Here, $M$ refers to the mass basis, which coincides with the $F$ basis in this context. In this case, $\left| f_{\nu_1^m}^{NO} - f_{\nu_2^m}^{NO} \right| = \left| X \right| = \frac{1}{N} \left| \sum_{k=1}^N (\beta_k^{ 2} - \gamma_k^{ 2}) \right|$ is, in principle, uniformly distributed and has only a $40\%$ chance of being less than 0.4. Moreover, $\left| f_{\nu_1^m}^{NO} - f_{\nu_2^m}^{NO} \right| \gtrsim 0.6$ occurs $40\%$ of the time in the context of Eq.~\eqref{rho-M}, whereas in the presence of $\mu$-$\tau$ mixing, it would be only $17\%$ (for $\sum\alpha_k^2 = 0$).

\item \textbf{Possible additional assumptions} \label{additional}

If we include additional assumptions$-$such as the improbability of extreme values of $\alpha_k^2$, or even $\beta_k^{\prime 2}$ and $\gamma_k^{\prime 2}$ in an arbitrary primed basis, which would imply that all neutrinos reaching the detector left the inner regions in the same state (an exception is made for the neutronization burst and early accretion phases, where $\sum \alpha_k^2 \approx 1$ is known to occur and collective effects are negligible.)$-$we can further narrow the range of $f_{\nu_e}^{NO}$, constraining it to the middle of the region described by Eq.~\eqref{NO5}, often resembling the value for IO in Eq.~\eqref{IO5}. 

\end{enumerate}

{\textbf {\textit {Antineutrinos.---}}}
For antineutrinos, the evolution Hamiltonian is similar to Eq.~\eqref{H} (main text), with the modification $V_e \rightarrow -V_e$. Therefore, at $\rho_m \gg \rho_m^H$, $\Bar{\nu}_e \approx \Bar{\nu}_1^m$ for NO, and the possible fraction of electron antineutrinos on Earth is given by Eq.~10 (main text) with $\left|U_{e 1}\right|^2$ in place of $\left|U_{e 3}\right|^2$,
\begin{equation} \label{AP1} \tag{A12}
    f_{\Bar{\nu}_e}^{NO}= \frac{1}{2} \left( 1 - \left|U_{e 1}\right|^2    \right) + \frac{a}{2} \left( 3 \left|U_{e 1}\right|^2 -1   \right).
\end{equation}
Assuming $\left|U_{e 1}\right|^2 \approx 2/3$ \cite{deSalas:2020pgw}, we have 
\begin{equation}  \label{AP2} \tag{A13}
    \frac{1}{6} \lesssim f_{\Bar{\nu}_e}^{NO} \lesssim \frac{2}{3}.
\end{equation}
Equation~\eqref{AP2} is applicable even in the presence of shock waves. On the other hand, for IO, $\Bar{\nu}_e \approx \Bar{\nu}_3^m$ above the $H$-resonance density, and $f_{\Bar{\nu}_e}^{IO}$ is described by Eq.~\eqref{NO5},
\begin{equation}  \label{AP3} \tag{A14}
    f_{\Bar{\nu}_e}^{IO} \lesssim 0.5.
\end{equation}
If the $H$-resonance is nonadiabatic, transitions $\Bar{\nu}_3^m \rightarrow \Bar{\nu}_1^m$ modify the allowed regions for $f_{\Bar{\nu}_e}^{IO}$, which will be given by Eq.~\eqref{AP1}. Therefore,
\begin{equation}  \label{AP4} \tag{A15}
    \frac{1}{6} \lesssim f_{\Bar{\nu}_e}^{IO} (\text{non-ad $H$}) \lesssim \frac{2}{3}.
\end{equation}
In case antineutrinos emerge from the core as incoherent matter eigenstates, we have $f_{\Bar{\nu}_e} \lesssim 0.7$ for both mass orderings.
These results are depicted in Fig.~\ref{fig2}.

\renewcommand{\thefigure}{A1}
\begin{figure*}[htb!] 
    \includegraphics[width=\textwidth]{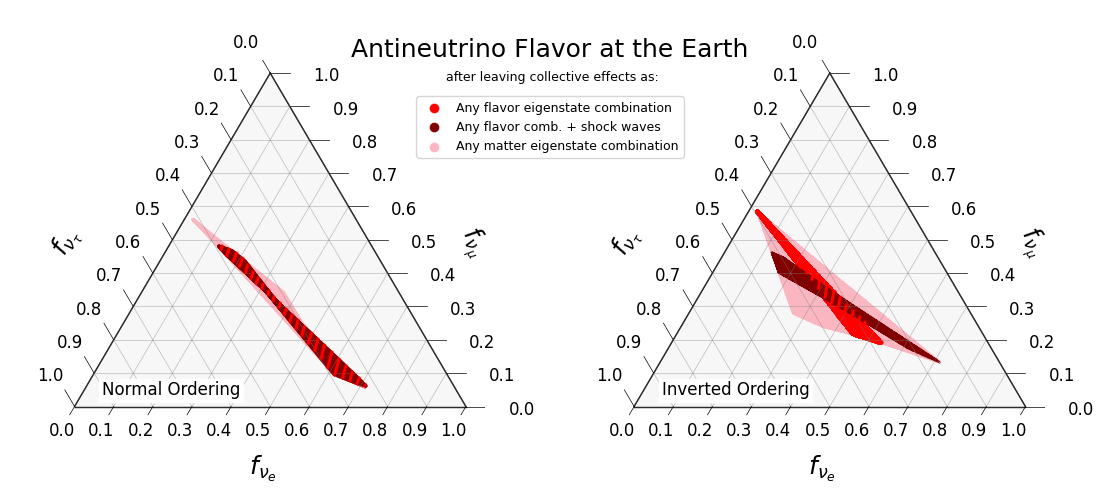}
    \caption{Same as Fig.~1 (in the main text), but for antineutrinos.}
    \label{fig2}
\end{figure*}

\vspace{0.05in}
 
\bibliographystyle{utcaps_mod}
\bibliography{reference}
\end{document}